# Robust Spin Interconnect with Isotropic Spin Dynamics in Chemical Vapour Deposited Graphene Layers and Boundaries

Dmitrii Khokhriakov[1], Bogdan Karpiak[1], Anamul Md. Hoque[1], Bing Zhao[1],

Subir Parui[2], Saroj P. Dash[1,3*]

[1]Department of Microtechnology and Nanoscience, Chalmers University of Technology, SE-41296, Göteborg, Sweden,
[2]K.U. Leuven, 3001 Leuven, Belgium,
[3]Graphene center, Chalmers University of Technology, SE-41296, Göteborg, Sweden.

**Abstract**

The utilization of large-area graphene grown by chemical vapour deposition (CVD) is crucial for the development of scalable spin interconnects in all-spin-based memory and logic circuits. However, the fundamental influence of the presence of multilayer graphene patches and their boundaries on spin dynamics has not been addressed yet, which is necessary for basic understanding and application of robust spin interconnects. Here, we report universal spin transport and dynamic properties in specially devised single layer, bi-layer, and tri-layer graphene channels and their layer boundaries and folds that are usually present in CVD graphene samples. We observe uniform spin lifetime with isotropic spin relaxation for spins with different orientations in graphene layers and their boundaries at room temperature. In all the inhomogeneous graphene channels, the spin lifetime anisotropy ratios for spins polarized out-of-plane and in-plane are measured to be close to unity. Our analysis shows the importance of both Elliott-Yafet and D'yakonov-Perel' mechanisms, with an increasing role of the latter mechanism in multilayer channels. These results of universal and isotropic spin transport on large-area inhomogeneous CVD graphene with multilayer patches and their boundaries and folds at room temperature prove its outstanding spin interconnect functionality, beneficial for the development of scalable spintronic circuits.

Spintronics aims at utilizing the electron spin degree of freedom for developing future data storage and processing devices, promising faster operation and lower power consumption.[1,2] So far, substantial progress has been achieved in non-volatile spintronic memory technology, where the spin-transfer torque and spin-orbit torque have enabled electrical control and switching of the magnetization.[3–5] For the realization of spin transistors and all-spin-logic devices, such non-volatile nanomagnetic elements should be connected by coherent spin channels with the possibility to manipulate the propagating spins.[2,6] Significant research efforts were devoted to finding a suitable spin interconnect exhibiting long spin lifetime and spin diffusion length, with spintronic studies of metals,[7] semiconductors,[8–11] and two-dimensional materials.[12–14]

Graphene has emerged as a promising interconnect material for spintronic applications due to its long spin diffusion length and low spin-orbit coupling.[14] Whereas the research on a single-device level is mainly done with the exfoliated graphene, large-scale fabrication prompts the use of a scalable production technique such as chemical vapor deposition (CVD). However, as opposed to a single-crystalline nature of exfoliated flakes, CVD growth can naturally lead to the formation of grain boundaries[15,16] and multilayer graphene patches.[17,18] For the successful integration of CVD graphene as a material of choice for the realization of long-distance spin interconnects in emerging spintronic devices, it is essential to investigate and understand the effect of these naturally-occurring inhomogeneities on spin relaxation properties. Particularly, one of the critical requirements for its application in spin interconnects is an experimental demonstration of universal and isotropic spin transport in CVD graphene, multilayer patches, and their boundaries at room temperature.

Theoretical studies of multilayer graphene and layer boundaries predicted their modified band structure, interlayer carrier hopping effects,[19,20] and the emergence of valley polarization,[21] edge states,[22] and interface Landau levels.[23] Experimentally, electrical transport properties of multilayer boundaries have been studied in epitaxial graphene[24,25] and, recently, in CVD-grown channels.[26] Owning to the transition from linear to parabolic energy dispersion at the layer boundary, a localized conductance minimum was observed due to poor wave function matching at the interface,[25] whereas the presence of edge-channel transport along the boundary was confirmed by quantum transport measurements.[26] In spintronics, theoretical investigations predicted that the spin relaxation by CVD-specific defects such as grain boundaries and ripples should not limit the device performance,[27] and experimental studies with continuous CVD graphene channels have confirmed their spin transport capabilities.[28–32] However, the spin scattering, dynamics, and spin lifetime anisotropy in multilayer CVD samples and their boundaries have not yet been determined experimentally.

Here, we demonstrate robust spin transport and isotropic relaxation in monolayer, bilayer, and trilayer graphene patches and across their layer boundaries that are naturally present in CVD grown graphene. The spin dynamics in continuous multilayer patches and through their boundaries and folds are probed by using spin valve and Hanle spin precession experiments at room temperature. The detailed measurements of spin precession in oblique magnetic fields and at various doping levels provide information about the spin dynamics, spin lifetime anisotropy and spin relaxation mechanisms. Our study suggests that inhomogeneities in large-area CVD graphene do not deter its spintronic device performance, showing universally isotropic spin transport properties at room temperature, which makes CVD graphene suitable for the realization of spin interconnects.

**Results and Discussion**

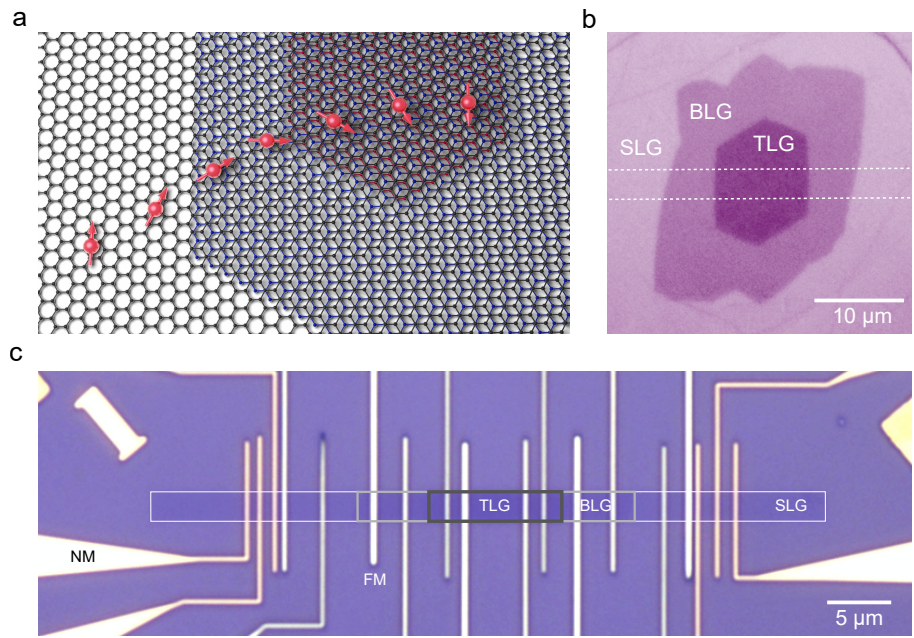

**Figure 1. A spintronic device with multilayer CVD graphene and layer boundaries. a,** A schematic of a multilayer graphene patch with diffusion and precession of spin-polarized electrons. **b**, An optical microscope image of a multilayer CVD graphene patch on SiO$_2$/Si substrate with an outlined area used for device fabrication spanning single-, bi-, and tri-layer graphene (SLG, BLG, TLG) areas, as well as the boundaries between them. **c**, A micrograph of a nanofabricated CVD graphene spintronic device used to investigate spin dynamics in SLG, BLG, TLG, and their layer boundaries with ferromagnetic (FM) tunnel contacts and nonmagnetic (NM) electrodes.

The graphene spintronic devices were specifically designed to investigate spin relaxation and dynamics in single-layer (SLG), bilayer (BLG), and trilayer (TLG) patches and across their layer boundaries. Figures 1a and 1b show a schematic and an optical picture of a multilayer CVD graphene patch transferred on a Si/SiO$_2$ substrate. The optical contrast allows to identify the SLG, BLG, and TLG regions. Depending on the CVD process parameters, adlayers of graphene have been shown to grow either underneath or on top of the first-formed layer, which is referred to as the Volmer–Weber (VW) and Stranski–Krastanov regimes, respectively.[33] In addition, the different growth regimes were utilized to produce Bernal-stacked[34] (twist angle 0°), turbostratic[35] (twist angle 30°), and arbitrary-twist-angle[36,37] graphene. Knowledge of the relative rotation angle is essential because if the graphene layers are not perfectly aligned, they form a moiré pattern, which alters its electronic properties and, for specific angles (*e.g.*, at a magic angle ~1.1°), gives rise to exotic electronic states including Mott insulators[38] and unconventional superconductors.[39] According to the information provided by the manufacturer (Grolltex Inc.), the graphene adlayers in our samples were formed beneath the initial layer, indicating the VW growth. As the corners of the BLG and TLG appear aligned,[40,41] and the VW regime predominantly yields Bernal-stacked graphene,[33] we consider that our patches have no twist angle and are not subjected to a moiré potential.

A CVD graphene device was patterned in a stripe shape spanning multilayer areas and their boundaries by electron beam lithography and oxygen plasma etching (Fig.1c). For the spintronic characterization of the devices, ferromagnetic tunnel contacts (Co/TiO$_2$) (FM) and nonmagnetic reference contacts (Au/Ti) were fabricated by subsequent EBL, metal evaporation and lift-off processes (see details in Methods).

## Spin transport and precession in multilayer graphene patches

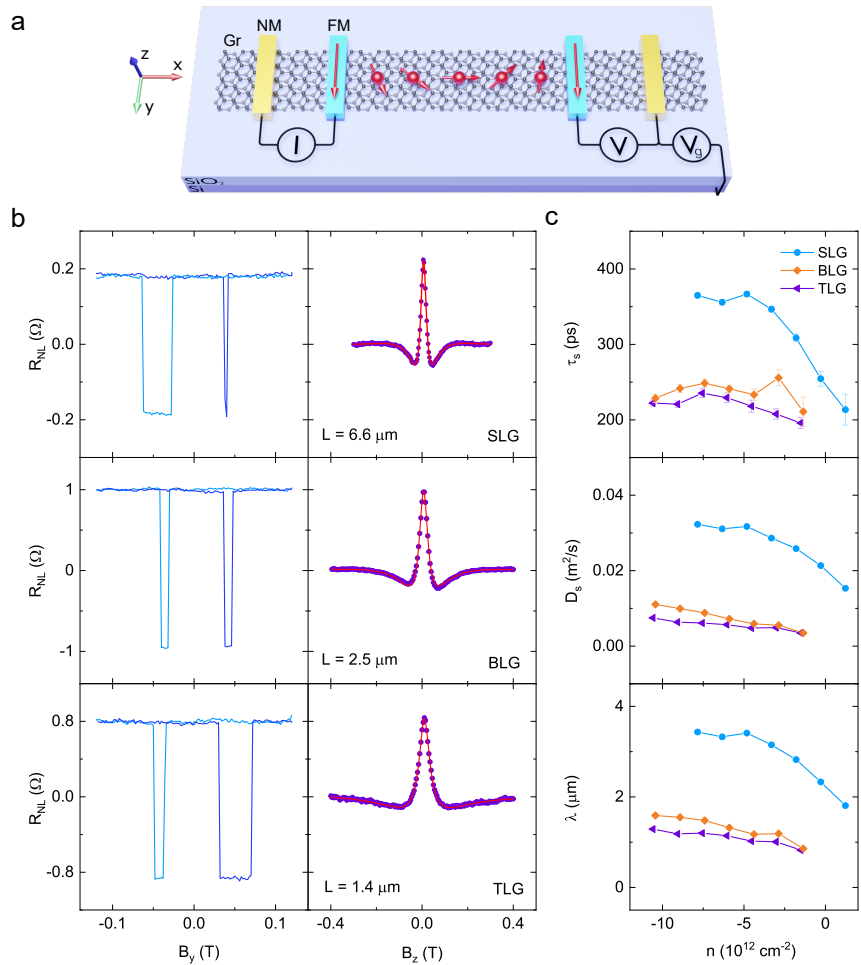

**Figure 2. Spin transport and precession in multilayer CVD graphene. a,** A schematic of the device and the nonlocal measurement geometry for spin transport and precession studies. **b,** The spin valve and Hanle spin precession signals measured in CVD SLG, BLG, and TLG channels. The channel lengths $L$ are noted in respective plots. The measurements were performed with $I = -50$ μA and $V_g = 0$ V. **c**, Spin lifetime, spin diffusion constant and spin diffusion length in SLG, BLG, and TLG extracted as a function of the carrier concentration ($n$) in graphene channels.

Spin transport in graphene was characterized using nonlocal (NL) spin valve (SV) and Hanle spin precession experiments at room temperature. The spin accumulation in graphene is created by passing a current $I$ between the injector ferromagnetic contact and graphene, while the NL spin-dependent voltage is detected by another ferromagnetic contact placed at a length $L$ away from the injector (Fig. 2a). To perform the spin valve measurement, we sweep the in-plane magnetic field $B_y$ along the easy axis of the ferromagnetic contacts while recording the nonlocal resistance. Sharp changes in $R_{NL}=V/I$ are measured when the magnetization of the injector or detector switches giving either parallel or antiparallel configuration, as shown in Figure 2b.

Hanle spin precession measurements were performed by measuring the nonlocal resistance while sweeping an out-of-plane magnetic field $B_z$, which induces spin precession and dephasing in the graphene channel. The data, shown in Fig. 2b, are fitted with eq. 1,

$$R_{\text{NL}}(B) = Re\left\{\frac{P^2 R_\square \widetilde{\lambda}_s}{W_{\text{gr}}} e^{-\frac{L}{\widetilde{\lambda}_s}}\right\} \quad (1)$$

with

$$\widetilde{\lambda}_s = \frac{\lambda_s}{\sqrt{1 + i\omega_L \tau_s}}$$

where $P$ is the average spin polarization of ferromagnetic contacts, $R_\square$ is the sheet resistivity of graphene, $W_{\text{gr}}$ is the graphene channel width, $\omega_L = \frac{g\mu_B}{\hbar}B$ is the Larmor spin precession frequency, $\mu_B$ is Bohr magneton, $L$ is the channel length, $\lambda_s = \sqrt{D_s \tau_s}$ is the spin diffusion length with $D_s$ and $\tau_s$ being the spin diffusion coefficient and spin lifetime, respectively. A small charge-based background signal was subtracted from the data by performing measurements with parallel and antiparallel configurations of the ferromagnetic electrodes (see Figure S1).

The corresponding SV and Hanle measurements in channels of SLG, BLG, and TLG are shown in Fig. 2b, establishing the presence of the spin transport and allowing to characterize spintronic parameters of the system. Performing such measurements at different gate voltages allows us to trace the parameters with changing carrier density in graphene, as shown in Fig. 2c. Strong *p*-type doping of our CVD samples (see Figure S2) restricts the accessible range of carrier concentration to mostly negative *n* values corresponding to hole conduction. We observe three main features in our data: (i) the spin parameters of SLG are higher than that of BLG and TLG, (ii) all parameters tend to reduce as the carrier density is reduced, (iii) the degree of tunability of the parameters reduces with the increasing number of layers.

The observed larger parameter values of SLG may indicate its superior capability for spin transport, but the difference can also appear due to the larger channel length of SLG compared to BLG and TLG.[29] Although the spin signal amplitude decreases exponentially with the channel length, long channels can be beneficial to reduce the adverse effects of additional defects and doping induced by the contacts into the underlying graphene,[42] which can diminish the spin parameters in shorter channels. In addition, whereas the presence of multiple graphene layers can be expected to protect the inner conducting paths from the scattering on magnetic surface impurities and increase the spin performance of multilayer graphene,[43,44] the interlayer hopping of carriers may introduce spin-dependent scattering effects.[45]

The reduction of spin transport parameters as the graphene Fermi level is tuned to the charge neutrality point (CNP) reflects the intrinsic relation between the carrier density and the diffusion constant, which also influences the spin lifetime due to the proportionality between $\tau_s$ and $D_s$ characteristic of the Elliot-Yafet spin relaxation mechanism.[46] In addition, the gate dependence studies can be affected by the conductivity mismatch between the contacts and graphene,[42] which in our devices leads to the reduction of spin signal magnitude near the CNP. However, as the resistance of our FM contacts ($R_c$ = 2 – 12 kΩ) is always larger than the channel spin resistance given by $R_s^{\text{gr}} = \frac{R_\square \lambda_s}{W_{gr}} \lesssim 1$ kΩ, the back–scattering of the spins into the FM electrodes should be suppressed and not have a strong influence on the obtained parameters. Finally, the smaller degree of tunability of the parameters with the carrier density in thicker channels is consistent with the notion of gate field screening, where the bottom layers of multilayer graphene are strongly coupled to the gate whereas the upper layers experience exponentially smaller field, and therefore are less tunable.[43]

# Spin dynamics in multilayer CVD graphene patches

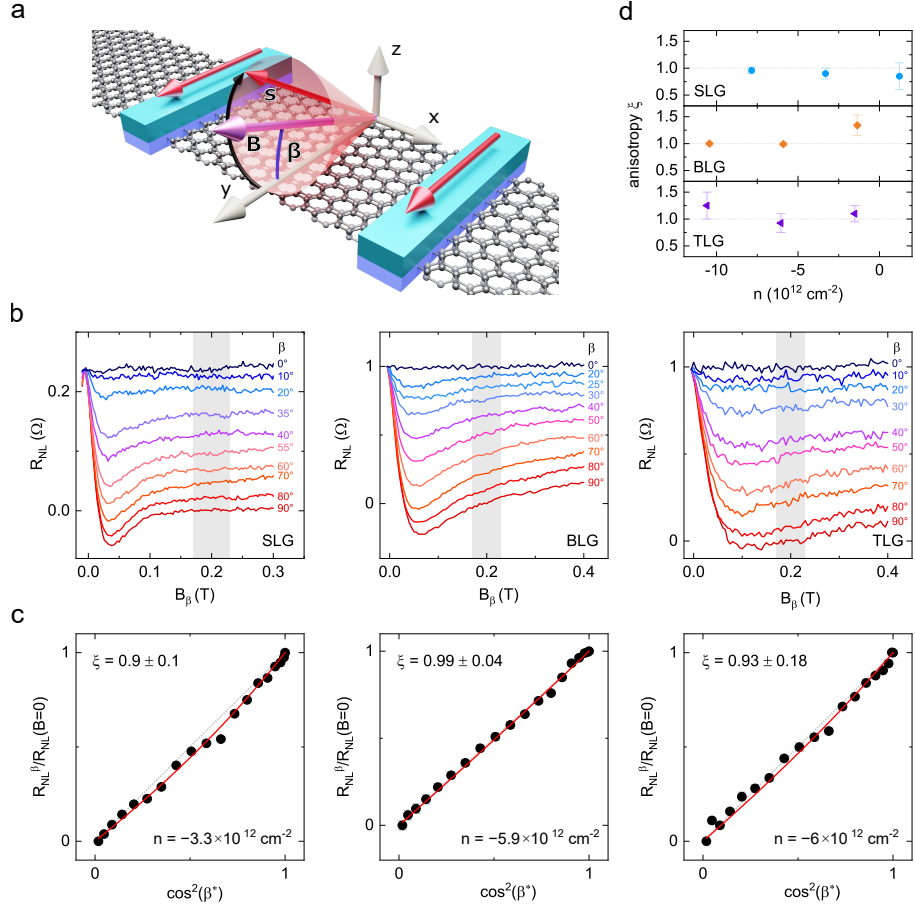

**Figure 3. Spin dynamics in multilayer CVD graphene. a,** Schematic of the spin precession in graphene with an oblique magnetic field. The field is applied at an angle $\beta$ to the FM contact easy axis (along $y$) in the $yz$-plane. **b,** Spin precession signals measured at different angles $\beta$ in SLG, BLG, and TLG at $V_g = 0$ V. **c,** Graphs visualizing the spin anisotropy, where the black dots represent the normalized data from respective panels in (b), the red line is a fit to eq. 2, and the dashed gray line is a guide to an eye for $\xi = 1$. **d**, Anisotropy values obtained in SLG, BLG, and TLG channels as a function of the carrier density. The measurements were performed with $I = -50$ µA at room temperature.

To get an insight into the nature of the spin dynamics in the CVD graphene multilayers, we study the spin lifetime anisotropy, which is defined as the ratio of relaxation times for spins polarized out-of-plane and in-plane $\xi = \frac{\tau_\perp}{\tau_\parallel}$. To quantify the spin anisotropy ratio, we employ spin precession in oblique magnetic fields.[47] In this method, the spin precession is studied with the external magnetic field applied at an angle $\beta$ to the contact magnetization in the $yz$-plane (Fig. 3a). Such measurement configuration leads to dephasing of spin components that are not parallel to the magnetic field. Thus, the resulting nonlocal resistance after complete dephasing is purely defined by the spin projection onto the field axis fixed by the angle $\beta$. Figure 3b shows the obtained spin precession signals for a set of $\beta$ in SLG, BLG, and TLG channels. To visualize $\xi$ in each channel, Fig. 3c shows the normalized value of the nonlocal resistance in the fully dephased regime $R_{NL}^\beta$ as a function of $\cos^2(\beta^*)$, where $\beta^*$ is the angle of the magnetic field corrected by a small angle $\gamma$ that describes an out-of-plane tilting of the injector and detector magnetization by the magnetic field. The values of $R_{NL}^\beta$ are obtained by averaging data points in a fully dephased regime marked by the shaded areas centered at $B_{sat}$ = 0.2 T to mitigate the measurement noise. Figure 3c allows to visualize the anisotropy, since for a completely isotropic system, the data would follow the $\xi$ = 1 line (dotted gray) whereas $\xi$ > 1(< 1) should render points above (below) that line. To obtain a quantitative estimation, the data are fitted with eq. 2 (red curves in Fig. 3c)[47]

$$\frac{R_{NL}^{\beta}}{R_{NL}^{0}} = \sqrt{\cos^2(\beta) + \frac{1}{\xi}\sin^2(\beta)}^{-1} \exp\left[-\frac{L}{\lambda_{\parallel}}\left(\sqrt{\cos^2(\beta) + \frac{1}{\xi}\sin^2(\beta)} - 1\right)\right] \cos^2(\beta^*) \qquad (2)$$

, where $L$ is the channel length, $\lambda_{\parallel} = \sqrt{D_s \tau_{\parallel}}$ is the in-plane spin diffusion length, $\beta^* = \beta - \gamma$, and $\gamma$ is the small angle of contact magnetization tilting out-of-plane between $B$ = 0 and $B_{sat}$. The extracted anisotropy values $\xi$ ~ 0.9 – 1 are close to unity, indicating mainly isotropic spin relaxation within the error margins of the experiment. Further, oblique spin precession measurements were repeated in three distinct carrier density regimes (low, medium and high density), giving a general overview of the system behavior as shown in Fig. 3d. The absence of a visible trend indicates that the CVD graphene does not develop noticeable directional spin-orbit fields in the studied carrier density regimes. This observation is similar to previous reports on exfoliated SLG, where only small and gate-independent anisotropy ($\xi$ ~ 0.8 – 1) was reported at both cryogenic and room temperatures.[47–49] Similarly, isotropic spin relaxation was previously observed in exfoliated BLG at room temperature, developing anisotropic and gate-dependent features at low temperatures.[50,51]

Whereas isotropic spin relaxation in multilayers of CVD graphene emphasizes its excellent performance as a spin interconnect material, a realization of active control over spin anisotropy in graphene can allow developing unique spintronic functionalities. To this end, gate-dependent anisotropy behavior may be possible to achieve in proximity-modified graphene by incorporating it in heterostructures with high-SOC TMDCs.[52,53] In addition, intriguing proximity-induced SOC effects have been reported in heterostructures of TMDCs and topological insulators with both exfoliated[54–58] and CVD graphene.[59–63]

# Spin transport and dynamics across layer boundaries in CVD graphene

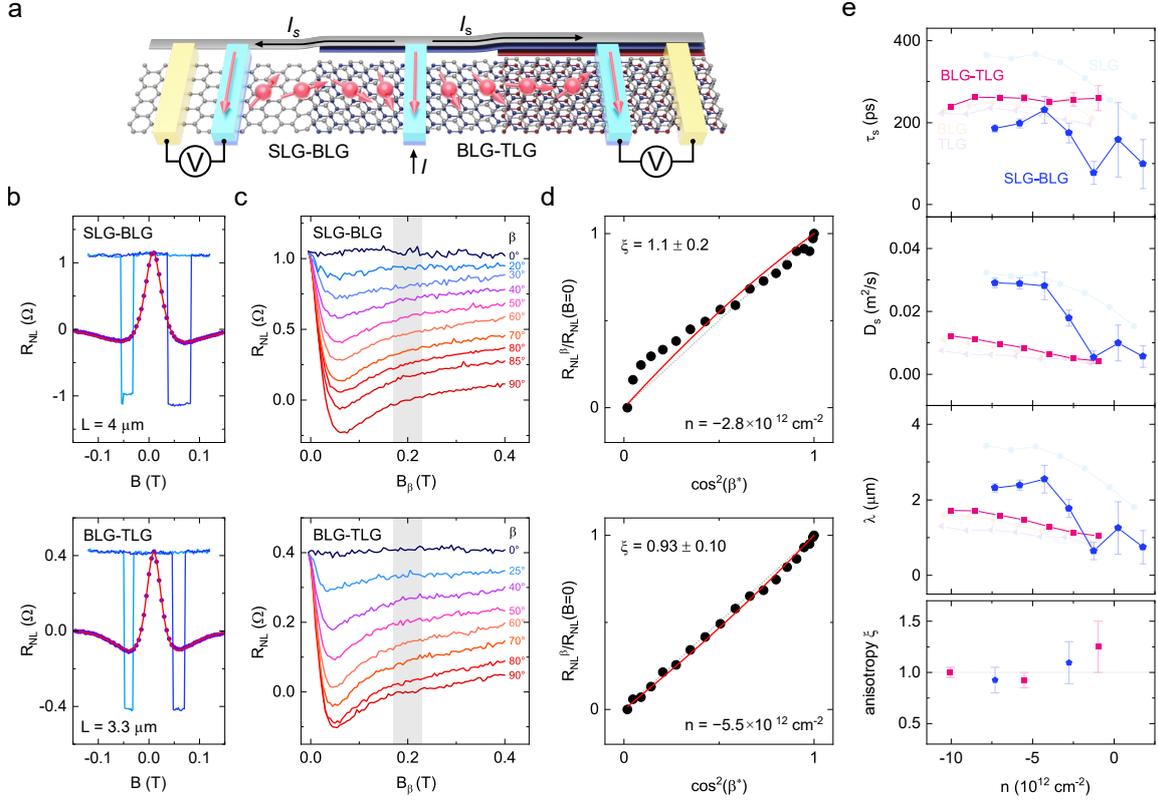

**Figure 4. Spin dynamics across the layer boundaries in CVD graphene. a,** A schematic of the device and measurement geometry to probe spin dynamics across the layer boundaries. **b,** The spin valve and Hanle signals in the channels with SLG-BLG and BLG-TLG boundaries. The data is obtained with $I = -50$ µA and $V_g = 0$ V. **c,** Spin precession signals across the SLG-BLG and BLG-TLG boundaries measured at different angles between the external magnetic field and contact magnetization in the $yz$-plane. **d**, Visualization of the spin anisotropy, where the black dots represent the normalized data from respective panels in (c), the red line is a fit to eq. 2, and the dashed gray line is a guide to an eye for $\xi$ = 1. **e**, Spin lifetime, spin diffusion constant, spin diffusion length, and spin lifetime anisotropy ratio $\xi$ across SLG-BLG (blue) and BLG-TLG (pink) boundaries as a function of the carrier concentration. Values for SLG, BLG, and TLG are shown in the background for comparison.

In addition to the presence of patches with different numbers of layers in CVD graphene, large-scale fabrication may result in channels incorporating layer boundaries. To study the effect of such boundaries on spin transport and spin lifetime anisotropy, we performed SV, Hanle, and oblique spin precession experiments in the channels having an SLG-BLG and BLG-TLG boundary (Fig. 4a). Figure 4b shows the obtained SV and Hanle signals across SLG-BLG and BLG-TLG boundaries, demonstrating their robust spintronic performance. Figure 4c shows the spin precession curves obtained with the magnetic field applied at different angles in SLG-BLG and BLG-TLG channels. The extracted anisotropy values (Fig. 4d) are close to unity, suggesting similar spin lifetime for spins polarized in and out of the graphene plane, which indicates that the interlayer hopping does not introduce anisotropic SOC-like scattering fields.

With an application of the gate voltage, we assess the dependence of spin transport parameters on the carrier concentration, as shown in Fig. 4e. The parameter values and their tunability across the BLG-TLG boundary are similar to those obtained in continuous BLG and TLG channels, whereas the SLG-BLG channel displays a larger range of values in its carrier density similarly to the SLG. A general reduction of $D_s$ and $\tau_s$ in thicker channels can be seen; however, the smaller values of $\tau_s$ in the SLG-BLG channel seem to deviate from the trend observed in other channels. It should be noted that $D_s$ and $\tau_s$ are independent spin transport parameters, affected differently by various aspects of device structure and fabrication processes. Previous studies of spin transport in graphene on various substrates showed that $D_s$ depends strongly on the substrate properties,[64,65] whereas $\tau_s$ is expected to be strongly affected by the channel contaminations and the presence of magnetic impurities. Thus, such behavior of $\tau_s$ in the SLG-BLG channel may be caused by the local variation of the density of spin scatterers. Spin lifetime anisotropy in channels with layer boundaries remains close to unity with an apparent upturn near the charge neutrality point, which, however, may be a consequence of a larger uncertainty in this regime due to the increase in the noise level of the measurements. In addition to the spin precession experiments with oblique magnetic fields, we used the xHanle measurement geometry, in which the spin precession in the $yz$-plane is induced by the magnetic field aligned in the $x$-direction ($B_x$). The measured curves could be fitted with the anisotropy parameter in the range of $\xi \sim 0.9 - 1.1$, showing a good agreement between the values obtained with the two techniques (see SI Note 1).

Spin transport in channels having a boundary between graphene areas of different thicknesses can help to identify the effects of interlayer diffusion of spin-polarized carriers and the related spin scattering timescale. As the graphene adlayers in our samples are grown beneath the continuous top layer, and all contacts are coupled to that upper layer, the detected spin transport could be weakly affected by the interlayer hopping. In experiments, the spin precession characteristics through layer boundaries do not show significant differences compared to the results obtained in continuous BLG and TLG channels. Comparing all studied channels, we observe predominantly isotropic spin scattering in all devices with similar spintronic performance, except for noticeably higher spin transport parameters in the SLG channel. Whereas this difference can stem from local variations in graphene properties and/or the more extended channels, it can also indicate that the interlayer spin scattering present in other channels leads to the reduction of spin transport characteristics.

## Spin dynamics across a fold in monolayer CVD graphene

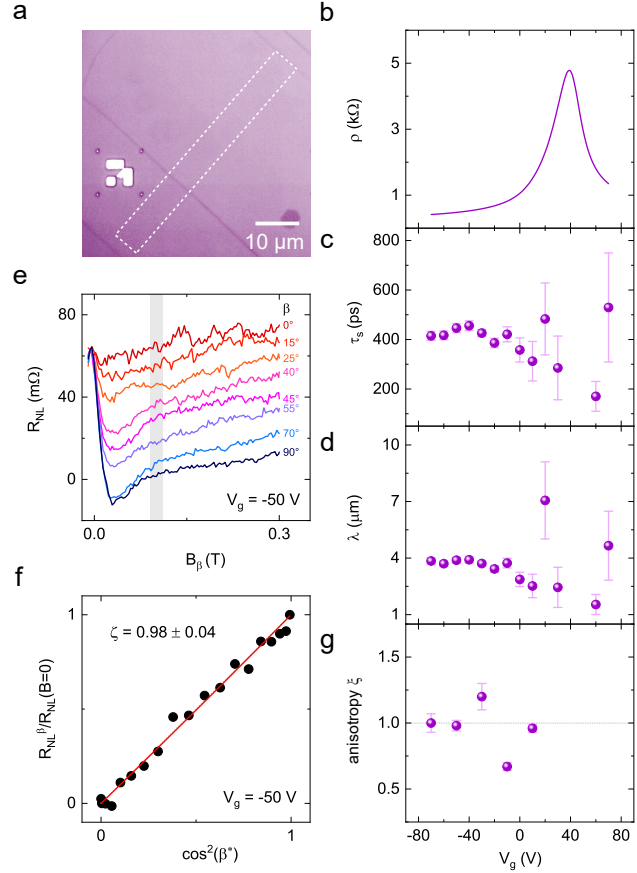

**Figure 5. Spin dynamics across a fold in monolayer CVD graphene. a,** An optical image of the monolayer CVD graphene with a fold. The device was patterned, as shown by the outline. **b,** The channel resistivity as a function of the back gate voltage, showing a Dirac point at $V_g$ = 39 V. **c,d,** The gate voltage dependence of the spin lifetime and spin diffusion length in the graphene channel with a fold. **e,** Hanle spin precession curves obtained for different angles of the applied magnetic field in the $yz$-plane. **f,** Normalized amplitude of the spin signal in a dephased regime plotted *vs* squared cosine of the field angle, allowing to visualize anisotropy. These measurements were performed in the channel with $L$ = 7.5 µm at $V_g$ = –50 V and $I$ = –300 µA. All the experiments were conducted at room temperature.

To investigate how the folds in a CVD graphene sheet can affect spin transport and spin lifetime anisotropy, we fabricated a device having such a fold in the channel, as outlined in Fig. 5a (see the final device picture in Fig. S4). The sheet resistivity of the channel is shown in Fig. 5b, indicating the Dirac point at $V_g$ = 39 V. A robust spin transport is observed in this channel at various gate voltages, with spin lifetime $\tau_s$ ~ 400 ps and spin diffusion length $\lambda$ ~ 4 µm. The obtained spin transport parameters are comparable with those extracted in the channels without irregularities, indicating that the presence of folds does not compromise the spin transport performance of wafer-scale CVD graphene.

To investigate the effects of the fold on spin lifetime anisotropy, we performed spin precession experiments in oblique magnetic fields, as shown in Figs. 5e,f. The obtained anisotropy values are close to unity and do not show any trend with the gate voltage (Fig. 5g), which indicates that the presence of a fold in graphene does not induce a noticeable anisotropic SOC at room temperature. The reduction of the signal-to-noise ratio due to the conductivity mismatch effect near the graphene Dirac point did not allow us to quantify the anisotropy for $V_g$ > 10 V reliably.

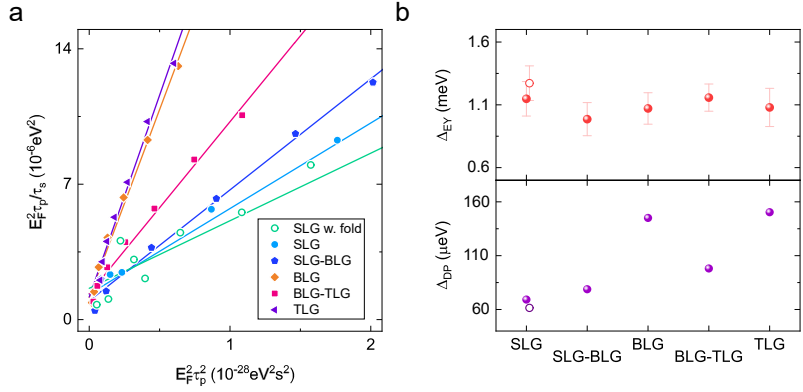

**Figure 6. Spin relaxation mechanisms in inhomogeneous CVD graphene channels. a,** Estimation of the contributions of spin scattering mechanisms in multilayer CVD graphene channels and their boundaries. The experimental data are fitted with eq. (3). **b,** The extracted SOC energy scale for the Elliott-Yafet (EY) and D'yakonov-Perel' (DP) mechanisms in different channels. Open circles correspond to the SLG channel with a fold.

**Spin scattering mechanisms**

To further investigate spin scattering in our inhomogeneous CVD graphene channels with multilayer patches, boundaries and folds, we consider the two main spin relaxation mechanisms. The Elliott-Yafet (EY) mechanism assumes spin-flip events during electron scattering, leading to the direct proportionality between spin lifetime $\tau_s$ and momentum scattering time $\tau_p$, whereas the D'yakonov-Perel' (DP) mechanism describes spin-flip scattering due to spin precession in effective magnetic fields between the scattering events, resulting in the spin lifetime being inversely proportional to the momentum scattering time. These mechanisms are associated with spin-orbit coupling (SOC) that induces band splitting, characterized by the energy scale parameters $\Delta_{DP}$ and $\Delta_{EY}$ with $\tau_{s,DP}^{-1} = 4\left(\frac{\Delta_{DP}}{\hbar}\right)^2 \tau_p$ and $\tau_{s,EY} = \left(\frac{E_F}{\Delta_{EY}}\right)^2 \tau_p$, where $E_F$ is the Fermi energy relative to the graphene Dirac point.[61,66] Assuming both mechanisms being relevant, we can define the total spin lifetime of the system $\tau_s = \left(\tau_{s,DP}^{-1} + \tau_{s,EY}^{-1}\right)^{-1}$, giving

$$E_F^2 \frac{\tau_p}{\tau_s} = \Delta_{EY}^2 + 4\left(\frac{\Delta_{DP} E_F \tau_p}{\hbar}\right)^2. \quad (3)$$

Figure 6a shows the experimental data for the studied channels fitted with eq. (3). The extracted $\Delta_{DP}$ and $\Delta_{EY}$ are shown in Fig. 6b, yielding comparable values with previous reports[67]. We observe similar $\Delta_{EY}$ values in different channels, whereas $\Delta_{DP}$ increases non-monotonically with increasing number of graphene layers.

Our results indicate an increased role of the DP mechanism in multilayer graphene, in agreement with previous studies.[28,44,68] However, other reports have found that the spin relaxation in both SLG and multilayer exfoliated graphene is limited by the EY mechanism.[43] Another theory has been proposed, assigning spin relaxation in both SLG and BLG to the resonant scattering on magnetic impurities, such as hydrogen adatoms and polymer residues.[69–71] This theory predicts opposite trends of the spin relaxation time in SLG and BLG near the CNP due to their different density of states and the scales of the energy fluctuations.

In SLG, the $\tau_s$ is predicted to increase with increasing carrier density, whereas in BLG it is predicted to have a nonmonotonic trend with a minimum at the intermediate doping and increasing $\tau_s$ towards both the CNP and strongly-doped regime. Our results in SLG CVD are consistent with this picture, however, all other inhomogeneous CVD channels with multilayer graphene patches, boundaries and folds display an apparent reduction of $\tau_s$ at large carrier concentrations, where the spin lifetime is limited by the D'yakonov-Perel' mechanism.

**Conclusion**

In conclusion, we demonstrate the robust spin interconnect performance of CVD graphene and spin dynamics in its characteristic defects such as multilayer patches, boundaries and folds. Our measurements show that these naturally occurring inhomogeneities in CVD graphene do not have a significant detrimental impact on its spin transport properties. The values of spin lifetime and spin diffusion length and the degree of their gate-tunability are similar in multilayer patches of different thickness. Using spin precession in oblique magnetic fields, we confirm mostly isotropic spin relaxation in all studied channels, where the anisotropy ratio is in the range of $\xi \sim 0.9 - 1.1$. The investigations of spin relaxation indicate the importance of both Elliott-Yafet and D'yakonov-Perel' mechanisms, with an increasing $\Delta_{DP}$ in multilayer patches. These experimental observations of robust spin transport in graphene multilayer patches and across their layer boundaries and folds at room temperature confirm the excellent performance of large-area CVD graphene as a spin interconnect material, with promising applications in emerging spin-based technologies and scalable spintronic circuits.

**Methods**

The CVD graphene spintronic devices on highly doped Si substrate with a thermally grown 285 nm $SiO_2$ layer were patterned by electron beam lithography (EBL) and oxygen plasma etching. The large area CVD graphene (obtained from Grolltex Inc) was grown on Cu foil and transferred onto a 4-inch $SiO_2$/Si wafer with pre-fabricated alignment markers. Both the nonmagnetic (Ti/Au) and ferromagnetic ($TiO_2$/Co) tunnel contacts were defined using EBL, followed by electron beam evaporation and lift-off processes. The ferromagnetic contacts were produced by electron beam evaporation of 6 Å of Ti in two steps, each followed by *in situ* oxidation in a pure oxygen atmosphere to form a $TiO_2$ tunnel barrier layer. In the same chamber, we deposited 40 nm of Co, after which the devices were finalized by lift-off in acetone at 65°C. The contact resistance was in the range of $R_c$ = 2 – 12 kΩ for FM electrodes, and $R_c$ = 2 – 5 kΩ for NM electrodes. In the measurements, a bias current was applied using a Keithley 6221 current source, and the nonlocal voltage was detected by a Keithley 2182A nanovoltmeter; the gate voltage was applied using a Keithley 2400 source meter.

**Acknowledgments**

Authors acknowledge the financial support from EU Graphene Flagship (Core 2 No. 785219 and Core 3 No. 881603), Swedish Research Council VR project grants (No. 2016-03658), 2D Tech Center (Vinnova), Graphene center, and the EI Nano and AOA Materials program at Chalmers University of Technology.

**Supporting Information**

The Supporting Information is available free of charge at https://pubs.acs.org/

Hanle spin precession with parallel and antiparallel FM configurations, conductivity and mobility data, spin precession with the in-plane field $B_x$, picture of a finished device with a graphene fold.

# Supporting information

# Robust Spin Interconnect with Isotropic Spin Dynamics in Chemical Vapour Deposited Graphene Layers and Boundaries

Dmitrii Khokhriakov[1], Bogdan Karpiak[1], Anamul Md. Hoque[1], Bing Zhao[1],

Subir Parui[2], Saroj P. Dash[1,3*]

[1]Department of Microtechnology and Nanoscience, Chalmers University of Technology, SE-41296, Göteborg, Sweden,
[2]K.U. Leuven, 3001 Leuven, Belgium,
[3]Graphene center, Chalmers University of Technology, SE-41296, Göteborg, Sweden.

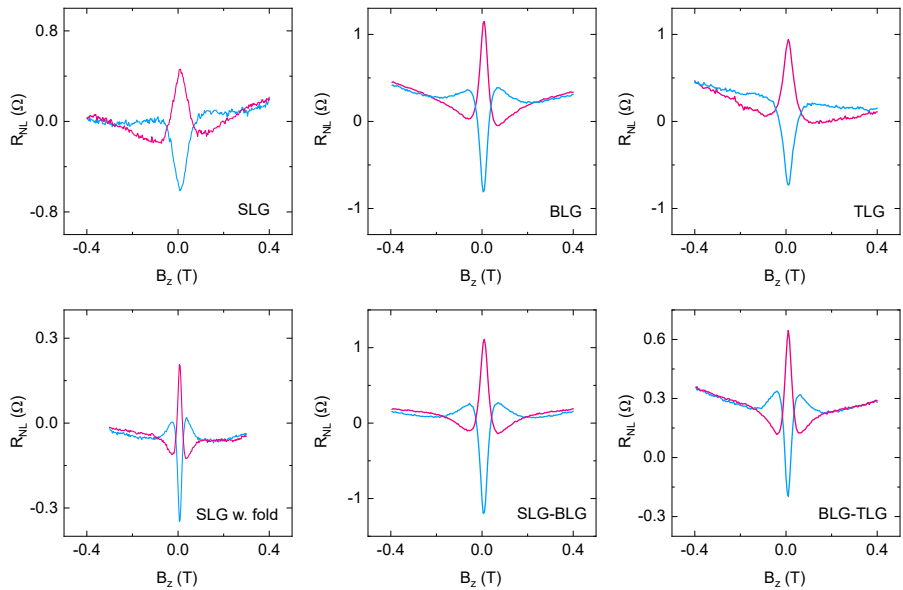

**Figure S1. Hanle spin precession in CVD graphene devices with multilayer patches and their layer boundaries with parallel and antiparallel FM configurations.** Hanle spin precession data for perpendicular magnetic field ($B_z$) in SLG, BLG, TLG channels and across SLG-BLG and BLG-TLG boundaries, obtained with parallel (pink curve) and antiparallel (blue curve) configuration of the ferromagnetic electrodes. By calculating the difference between the two curves, the parabolic charge-based background signal was removed in Figs. 2b and 4b. For SLG channel described in the main text, the background was subtracted manually; the first panel here shows representative data from another SLG device. All the measurements were performed at room temperature.

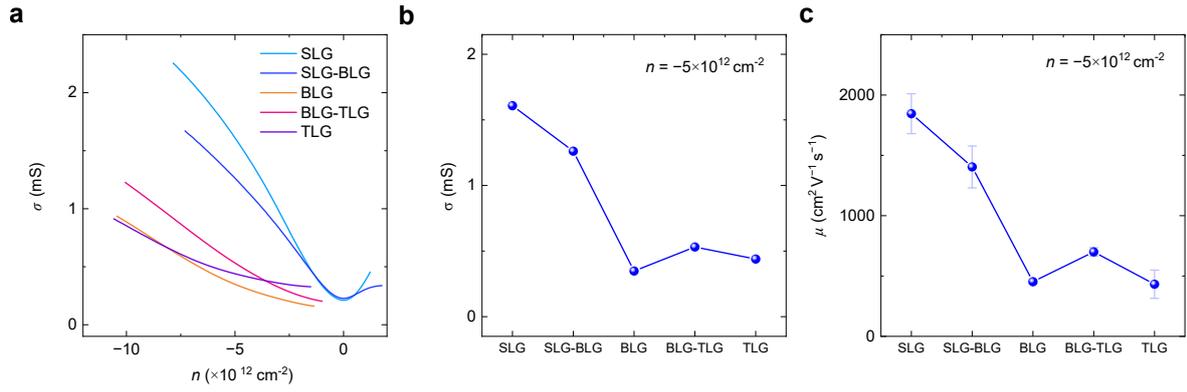

**Figure S2. Conductivity and mobility in multilayer graphene channels and across their boundaries. a**, Conductivity of the studied channels as a function of the carrier density. **b,c**, Conductivity and mobility extracted at n = –5×10$^{12}$ cm$^{-2}$ in different channels. The graphs indicate a general trend of reducing mobility and conductivity with increasing number of graphene layers. This trend in electrical performance of multilayer channels compared to SLG is consistent with sizeable interlayer coupling expected for Bernal graphene.[1] The mobility of our samples has typical values for wet-transfer CVD graphene, smaller than mobilties commonly observed in exfoliated devices. However, previous studies of spin transport in samples with different mobility,[2,3] as well as in high-mobility exfoliated graphene on hBN,[4] did not show a significant correlation between $\tau_s$ and $\mu$. Thus, we expect our findings to be valid also for the exfoliated graphene samples of various mobilities.

# Note 1

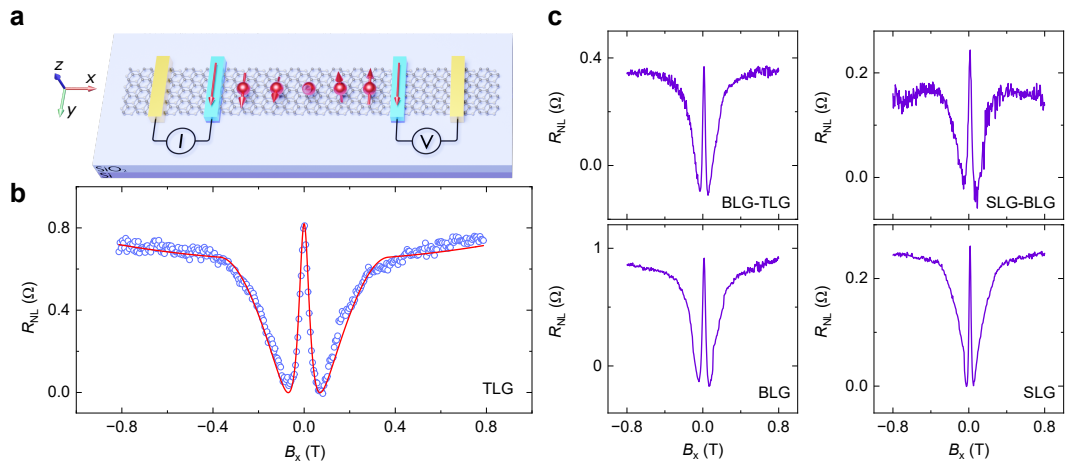

**Figure S3. Spin precession in CVD graphene devices with multilayer patches and their layer boundaries with B$_x$ field sweep. a**, A schematic of the device with measurement geometry and spin precession in $B_x$. **b**, The measured xHanle signal in a TLG channel (data points) with fitting (red line). **c**, xHanle data obtained in SLG and BLG channels, as well as in the channels having SLG-BLG and BLG-TLG boundary. All the measurements were performed at room temperature.

The spin precession was also measured in all channels while sweeping the magnetic field $B_x$ in the graphene plane along the channel (Fig. S3a), yielding the xHanle curves presented in Fig. S3b,c. These curves show a combination of spin precession in the $yz$ plane and rotation of injector and detector polarization in the $yx$ plane, which leads to the contacts aligning with the filed in $x$-direction above the saturation field $B_{\text{sat}} \approx 0.4$ T. Due to the spin precession happening in the $yz$ plane, both in-plane and out-of-plane spin lifetimes influence the measurement, and therefore this geometry allows to quantify the spin lifetime anisotropy ratio $\xi$. Since the anisotropy information is fully contained in the spin precession contribution, it is necessary to isolate it by removing from the signal the additional contribution originating from magnetization rotation. Normally, this contribution can be completely subtracted by performing xHanle measurements with parallel and antiparallel initial configurations of the ferromagnetic contacts, since in such measurements the spin precession signal changes sign whereas the magnetization rotation does not. Unfortunately, our device has failed before the measurements in the AP configuration could be performed. In addition, the two contributions can be easily separated in long channels, in which complete spin dephasing can be achieved already in small fields; however, our channel lengths were limited by the device geometry and size of the multilayer patch. Therefore, we could simulate the magnetization rotation contribution in our device using the Wohlfarth-Stoner model, but the saturation level of our signal at high fields differs from the signal value at $B = 0$, which hinders data interpretation since at both $B = 0$ and $|B| > B_{\text{sat}}$ the same value of spin signal is expected due to ferromagnetic contacts being in the parallel configuration ($\boldsymbol{M_i} \parallel \boldsymbol{M_d} \parallel y$ and $\boldsymbol{M_i} \parallel \boldsymbol{M_d} \parallel x$ respectively). In our device, the application of the magnetic field gives rise to an additional small parabolic background signal due to the graphene magnetoresistance; however, this magnetoresistance is positive and, therefore, cannot be the reason for the lower signal at high fields. Alternatively, this difference in the signal magnitudes may indicate varying spin lifetimes for spins aligned in $x$ and $y$ directions, but such anisotropy between the in-plane directions is not expected due to the symmetry of the graphene lattice. Thus, the origin of such missing signal is still unidentified, although this feature has also been observed in exfoliated single-layer graphene devices.[5]

Using a spin precession model that includes both anisotropic spin relaxation and contact magnetization rotation,[6,7] we fit the data, as shown in Fig. S3b. To account for the missing part of the signal in our fitting, we introduce an additional scaling factor, which, however, limits the accuracy of parameter extraction. Within the noise level of our experiment, all data can be adequately fitted with $\xi$ being in the range of 0.9 – 1.1, indicating isotropic spin relaxation, in agreement with the oblique Hanle spin precession experiments described in the main manuscript.

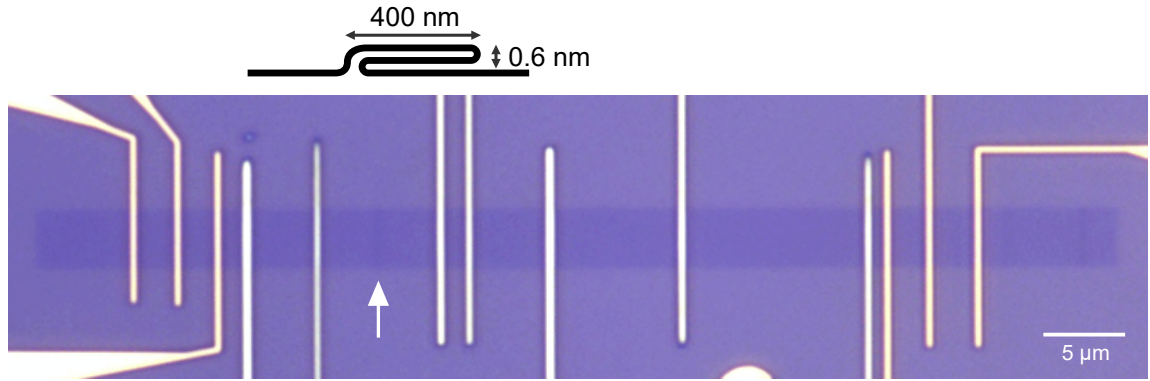

**Figure S4. Optical microscopy photograph of the CVD graphene device with a fold.** A micrograph of a nanofabricated CVD graphene spintronic device used to investigate spin dynamics across the fold in monolayer graphene. The arrow points at the fold in the graphene channel. The atomic force microscopy was used to extract the dimensions of the fold, which are shown in the schematic.